\documentclass[aps,prl,showkeys,showpacs,groupedaddress,amsmath,amssymb,floatfix,twocolumn]{revtex4-1}

\usepackage{pdfpages}
\usepackage{color} 
\usepackage{todonotes}

\usepackage{lmodern}						

\begin{document}


\title{Traveling phase waves in asymmetric networks of noisy chaotic attractors}
\author{Thomas K. DM. Peron$^{1,2}$}
\email{thomaskaue@gmail.com}
\author{J\"urgen Kurths$^{2,3}$}
\author{Francisco A. Rodrigues$^{4}$}
\author{Lutz Schimansky-Geier$^{3}$}
\author{Bernard Sonnenschein$^{3}$}
\email{sonne@physik.hu-berlin.de}

\affiliation{$^{1}$Instituto de F\'isica de S\~ao Carlos, Universidade de S\~ao Paulo, CP 369, 13560-970 S\~ao Carlos, S\~ao Paulo, Brazil}
\affiliation{$^{2}$Potsdam Institute for Climate Impact Research (PIK), 14473 Potsdam, Germany}
\affiliation{$^{3}$Department of Physics, Humboldt-Universit\"at zu Berlin, Newtonstrasse 15, 12489 Berlin, Germany}
\affiliation{$^{4}$Instituto de Ci\^encias Matem\'aticas e de Computa\c{c}\~ao, Universidade de S\~ao Paulo, CP 668, 13560-970 S\~ao Carlos, S\~ao Paulo, Brazil}


\begin{abstract}
We explore identical R\"ossler systems organized into two equally-sized groups, among which differing positive and negative in- and out-coupling strengths are allowed. Patterns of distinctly synchronized phase dynamics are observed, which coexist with chaotically evolving amplitudes. In particular, we report the emergence of traveling phase waves, i.e. states in which the oscillators settle on a new rhythm different from their own. We further elucidate our findings through phase-coupled R\"ossler systems, establishing a connection with the Kuramoto model. Together with the study of noise effects, our results suggest a promising new avenue towards the coexistence of chaotic, noisy and regular collective dynamics.
\end{abstract}

\maketitle 

\textit{Introduction.}--- Discovering that coupled chaotic oscillators can synchronize their phase angles \cite{rosenblum1996phase,*rosenblum1997FromPhaseLagSync}, marked a milestone in the study of collective synchronization. The fascination was due to the fact that the synchronous behavior was largely hidden, because the amplitudes of the oscillators remained chaotic and uncorrelated. This phenomenon has now found general interest in many practical applications \cite{osipov2007synchronization}.

In this Letter we show that even more subtle collective rhythms are possible among chaotic oscillators. It is demonstrated that mixed attractive-repulsive couplings among identical chaotic oscillators yield \textit{traveling phase waves} (TW) and $\pi$-\textit{states}. In the former the oscillators agree on a new rhythm different from their own, i.e. the locking frequency does not equal the natural frequency, while in the $\pi$-state the oscillators split into two clusters separated by a phase lag of $\pi$. In all these states the oscillators remain chaotic and uncorrelated in their amplitudes.

The existence of TWs has remained unnoticed in the
phase synchronization of chaotic systems. Our work provides a significant broadening for the emergence of those collective phenomena, encompassing chaotic oscillatory behavior with coupled phase-amplitude dynamics. In particular, we investigate two different formulations for the coupling setup between the oscillators. 
Firstly, in the \textit{weighted $x$-coupled model}, we consider R\"ossler oscillators~\cite{rossler1976equation} with funnel and phase-coherent attractors coupled through the $x$ coordinate. In this case, besides the occurrence of $\pi$-states, a novel dynamical state is reported. Namely, in contrast to oscillatory systems~\cite{HoStr11PRL,IaPeMcSte13,sonnenschein2015collective}, it is found that a noticeable spontaneous drift in the frequencies occurs only when the system is in the incoherent state. These properties define the \textit{incoherent TW state}. Secondly, in the \textit{phase-coupled model}, oscillators operating 
in the phase-coherent regime interact purely through the sine of the 
difference of their phases. Even though the phase angles depend on the chaotic amplitudes, our findings bridge the gap to the paradigmatic Kuramoto model,  providing at the same time a rounded picture on the emergence of traveling phase waves.
\begin{figure}[!tpb]
\begin{center}
\includegraphics[width=1.0\linewidth]{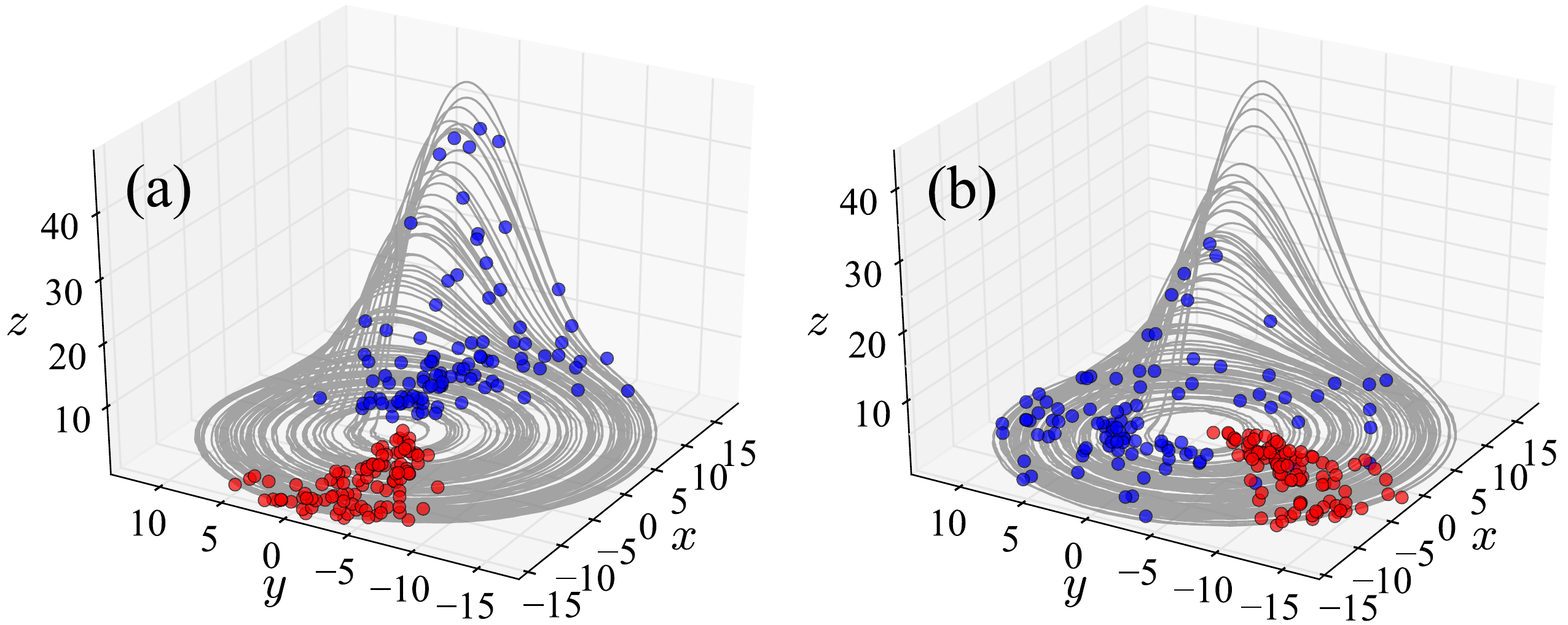}
\end{center}
\caption{Snapshots of the oscillators' trajectories of the weighted $x$-coupled R\"ossler model (Eq.~\ref{eq:rossler_x_coupling}) with phase-coherent attractor ($a=0.15$, $b=0.2$ and $c=10$). (a) $\pi$-state where $\delta = \pi$ with $K_0 = 0$ and $G_0 = 0$. (b) Lag sync state for which $\delta < \pi$ with $K_0 = 0$ and $G_0 = 0.2$. Gray line in the background corresponds to a trajectory of a randomly selected oscillator in order to depict typical dynamics. Remaining  parameters: $N = 200$, $\Delta K = 0.1$, $\Delta G = 1$ and $D = 0.03$. See~\cite{SM} for the animated version of the figure and additional videos on other collective states.}
\label{Fig1}
\end{figure}

\textit{R\"ossler systems with diffusive coupling.}---
Let us consider first a coupling setup where the oscillators interact through the $x$ coordinate via a weighted network~\cite{shima2004rotating,*gu2013spiral,gomez2013diffusion,*boccaletti2014structure}
 defined by the matrix $\mathbf{W}= [w_{ij}] $ as follows
\begin{equation}
\begin{aligned}
\dot{x}_{i}= & -\omega_{0}y_{i}-z_{i}+\frac{1}{N}\sum_{j=1}^{N}w_{ij}(x_{j}-x_{i})+\xi_{i}(t),\\
\dot{y}_{i}= & \ \omega_{0}x_{i}+ay_{i},\ \ \dot{z}_{i}=b+z_{i}(x_{i}-c).
\end{aligned}
\label{eq:rossler_x_coupling}
\end{equation}
The phases can be defined as 
$
\phi_i^{\rm{H}}(t)=\arctan\left[x^{\rm{H}}_i(t)/x_i(t)\right], 
$
with $x^{\rm{H}}_i$ being the Hilbert transform of $x_i$,
$
x^{\rm{H}}_i(t) = \pi^{-1} \int_{-\infty}^{\infty} x_i(\tau)/\left(t-\tau\right) d\tau,
$ where the integral is taken in the sense of the Cauchy principal value~\cite{freund2003frequency}. The terms $\xi_i(t)$ correspond to time-dependent disorder modeled as Gaussian white noise satisfying 
$
\left\langle \xi_{i}(t)\right\rangle = 0,\ \left\langle \xi_{i}(t)\xi_{j}(t)\right\rangle = 2D\delta_{ij}\delta(t-t'), 
$
where $D$ is the noise strength.

The coupling between the oscillators is mediated through the weights $w_{ij}$, which are henceforth given by $w_{ij} = K_i G_j$. In effect, to each oscillator a pair of couplings $(K_i,G_i)$ is associated. Coupling strength $K_i$ is responsible for how strongly node $i$ responds to the interaction with the rest of the population, whereas 
$G_i$ accounts for how strongly it influences the dynamics of other oscillators. Furthermore, in this formulation, the coupling scheme in Eq.~\ref{eq:rossler_x_coupling} can be seen as a linear diffusive process between different groups with negligible spatially embedding effects~\cite{shima2004rotating,*gu2013spiral,gomez2013diffusion,*boccaletti2014structure} 

Heterogeneity in the interaction patterns is introduced by dividing the population of $N$ oscillators into two equally-sized subpopulations, denoted by labels ``1'' and ``2'', and characterized by two pairs of couplings $(K_1,G_1)$ and $(K_2,G_2)$. All these constants are allowed to be positive or negative.  Henceforth, we adopt the parametrization
\begin{equation}
K_{1,2} = K_0 \pm \frac{\Delta K }{2} \mbox{ and } G_{1,2} = G_0 \pm \frac{\Delta G}{2}, 
\label{eq:parametrization}
\end{equation}
where $K_0$, $G_0$ are the average in- and out-coupling and $\Delta K$, $\Delta G$ the respective mismatches. Furthermore, we speak of mixed interactions, if both positive and negative couplings are present, i.e., always when the condition $|\Delta K|/2>K_0$ or $|\Delta G|/2>G_0$ is satisfied. 

Having defined the phases, the level of phase synchronization is measured by the order parameter 
$
r = \left\langle \left| N^{-1}\sum_{j=1}^N \exp\left[i\phi^{\rm{H}}_j(t)\right] \right|\right\rangle_t, 
$
where $|\cdot|$ is the absolute value and $\left\langle \cdots \right\rangle_t$ denotes temporal 
average. Organising the index labelling such that $j \in \{1,...,N/2\}$ refers to oscillators belonging to subpopulation 1 and $j \in \{N/2+1,...,N\}$ to subpopulation 2, it is also
convenient to calculate the coherence within each group as 
$ r_{1}e^{i\Theta_{1}(t)} = (N/2)^{-1} \sum_{j=1}^{N/2} \exp [i\phi^{\rm{H}}_j(t)]$ and analogously for $r_2e^{i\Theta_2(t)}$, with $\Theta_{1,2}(t)$ being the corresponding mean phases. 
The phase lag $\delta(t)$ is then defined
as the difference between the mean phases  $\delta(t) = |
\Theta_1(t) - \Theta_2(t)|$. Moreover, we calculate the 
wave speed as
\begin{equation}
\Omega = \frac{1}{N}\sum_{j=1}^N \left\langle \frac{d\phi^{\rm{H}}_i}{dt} \right\rangle_t.
\label{eq:wave_speed}
\end{equation} 

\begin{figure}[!tpb]
\begin{center}
\includegraphics[width=1.0\linewidth]{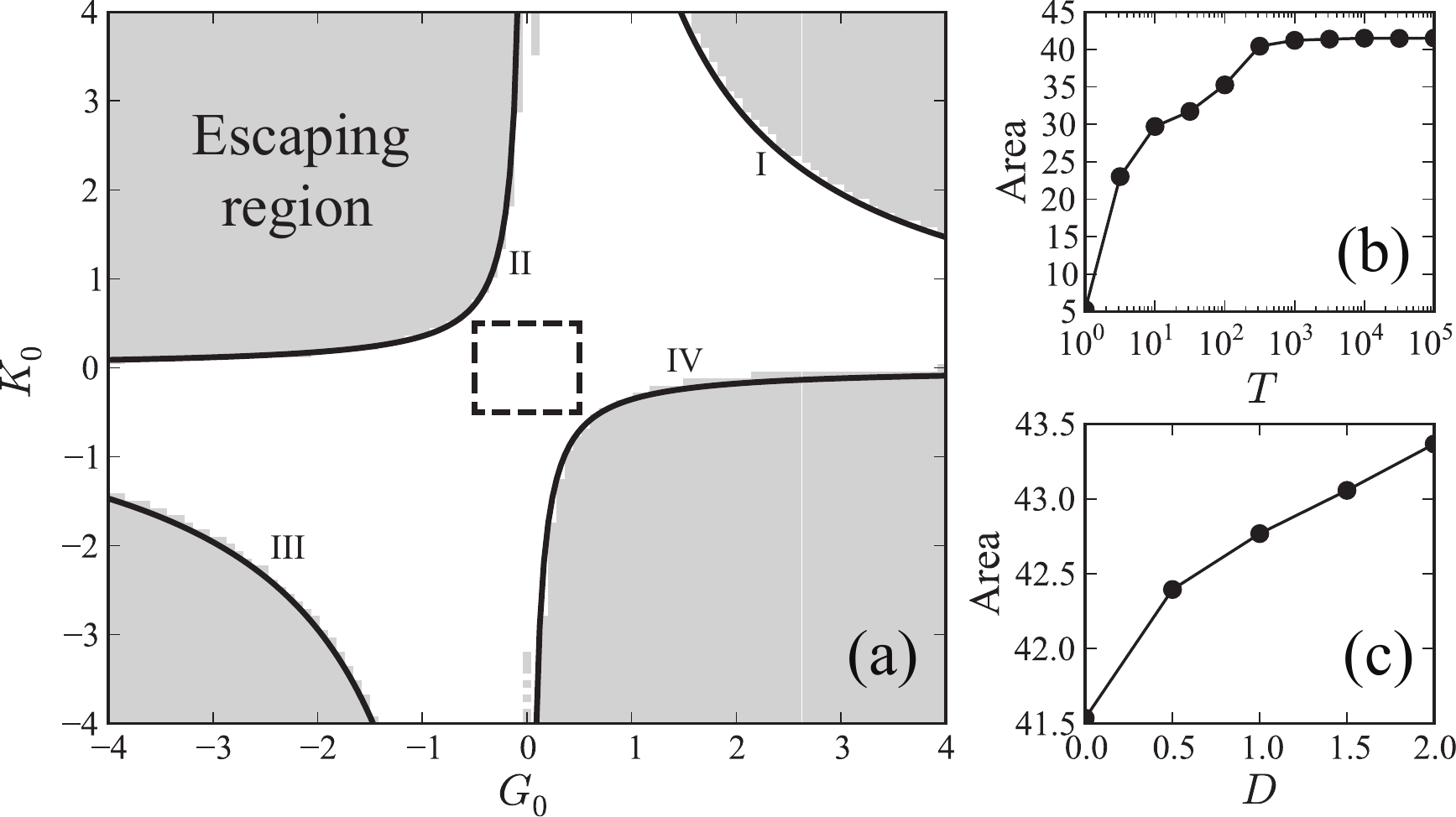}
\end{center}
\caption{ (a) Map showing the regions with escaping trajectories for weighted $x$-coupled R\"ossler oscillators (Eq.~\ref{eq:rossler_x_coupling}) in the phase-coherent regime  considering $N = 500$, $D=0$, $\Delta K =0.1$ and $\Delta G = 1$. For each pair $(G_0,K_0)$ the initial conditions $[x_i(0),y_i(0),z_i(0)]$ $\forall i$ are randomly drawn according to a uniform distribution in the range $[-1,1]$. Simulation time $T = 1\times 10^5$. Solid black curves fitting the boundaries of the escaping region are obtained through the relation $\langle \langle KG \rangle \rangle = c$, where $c_{\rm{I}} = c_{\rm{III}} = 11.8$ and $c_{\rm{II}} = c_{\rm{IV}} = -0.66$. Square at the center depicts the parameter region of Fig.~\ref{Fig3}. Area of escaping region as function of (b) simulation time $T$ for fixed $D=0$ and (c) as function of $D$ for $T = 5 \times 10^3$.}
\label{Fig2}
\end{figure}

In terms of the collective
variables the dynamical states are expressed as follows: (i)
In the incoherent state, oscillators of both populations 
rotate independently so that $r_{1,2}\approx 0$. (ii) The 
oscillators are mutually attracted in the partial phase synchronized
state yielding $r>0$ and $\delta = 0$ which we denominate as ``zero-lag sync''. (iii) A similar state to (ii) is attainable by system (\ref{eq:rossler_x_coupling}). Specifically, we refer to ``lag sync'' when partial global synchronization is achieved ($r>0$) with $0 < \delta < \pi$ and in the absence of additional drift ($| \Omega - \omega_0| \simeq 0$). (iv) In the $\pi$-state oscillators belonging to the same subpopulations 
are partially synchronized in phase ($r_{1,2}>0$), while the mean 
phases are separated by $\delta = \pi$. (v) Finally, in the TW 
state, the phases drift with a frequency different from their 
intrinsic one yielding $|\Omega - \omega_0|>0$. 
Figure~\ref{Fig1} depicts snapshots of the trajectories of $N = 
200$ weighted $x$-coupled R\"ossler oscillators 
(Eq.~\ref{eq:rossler_x_coupling}). Precisely, Fig.~\ref{Fig1}(a) 
shows a typical long-term configuration of a $\pi$-state, where 
the populations are diametrically opposed. A typical lag sync 
state in the weighted $x$-coupled R\"ossler model is exemplified 
in Fig.~\ref{Fig1}(b), where the centroids of the two clusters 
are separated by a phase-lag $\delta < \pi$. We shall see that 
the TW states are manifested in particular distinctive ways 
depending on the coupling model.

Before we systematically investigate the dynamical states exhibited by model (\ref{eq:rossler_x_coupling}), we should remark on the possibility of facing trajectories that escape the attractor after a period of transient chaos. Divergent trajectories arise due to the dissipative character of the R\"ossler dynamics and are in general related to the conversion of a chaotic attractor into a nonattracting chaotic invariant set~\cite{letellier1995unstable,*barrio2014UnboundedDynamics}. In order to show how this is manifested in weighted $x$-coupled R\"ossler oscillators in the absence of noise, Fig.~\ref{Fig2} depicts a sample of the plane $G_0 \times K_0$ where trajectories that diverge to infinity are possible. As it is seen, for large values of $|G_0 K_0|$, divergences from the attractor are inevitable, but still a wide region with non-divergent orbits is observed. Noteworthy, the collective escapes are induced by the coupling, since isolated oscillators do not escape for the parameters considered here as shown in~\cite{letellier1995unstable,*barrio2014UnboundedDynamics}.

\begin{figure}[!tpb]
\begin{center}
\includegraphics[width=1.0\linewidth]{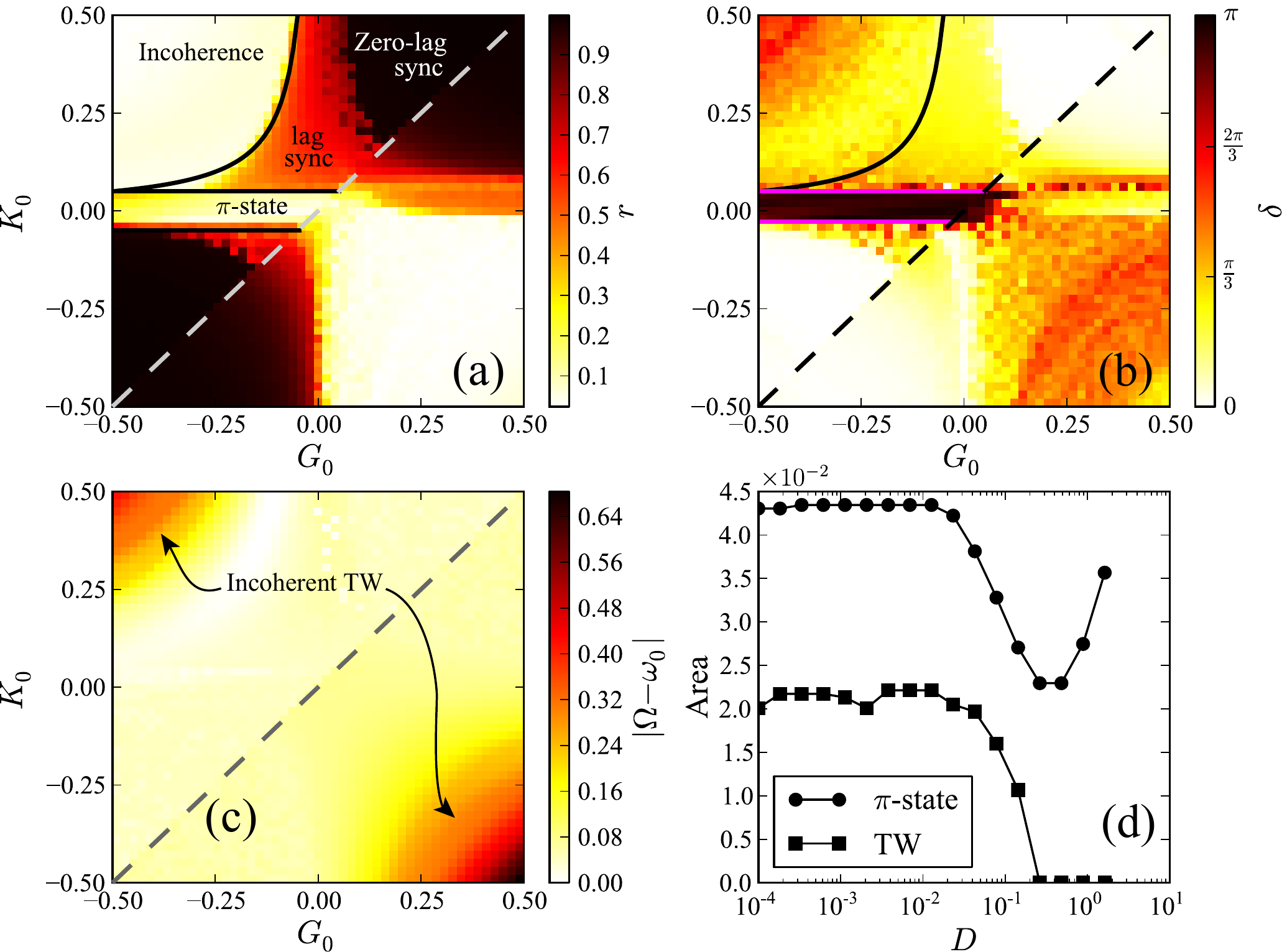}
\end{center}
\caption{Colormaps showing (a) order parameter $r$, (b) phase lag $\delta$ and (c) speed of phase wave $|\Omega - \omega_0|$ for weighted $x$-coupled R\"ossler oscillators (Eq.~\ref{eq:rossler_x_coupling}) with phase-coherent (upper triangles) and funnel (lower triangles) attractors considering $D=0.03$. (d) Area of  TW state in which $|\Omega - \omega_0| \geq \varepsilon$, where we consider $\varepsilon = 0.3$; and $\pi$-state for $\delta \geq \varepsilon$ with $\varepsilon = 3$. Remaining parameters: $\Delta K = 0.1$, $\Delta G = 1$, $N = 10^3$, $50 \times 50$ grid, simulation time $T = 5\times 10^3$. Curves in the interface between incoherent and partial synchronized state for phase-coherent oscillators are given by the condition $2D_{\textrm{eff}} = \langle \langle KG \rangle\rangle$, here with
 $D_{\textrm{eff}} \approx 0$~\cite{Deff}. Horizontal lines correspond to the critical points $K_{0c} = \pm \Delta K/2$. }
\label{Fig3}
\end{figure}

Taking Fig.~\ref{Fig2} as a guide so that the parameters are chosen in a region free 
from escaping trajectories, in Fig.~\ref{Fig3} we show the dependence of $r$, $\Omega$ 
and $\delta$ as a function of the mean in- and out-couplings $K_0$ and $G_0$, considering oscillators with phase-coherent ($(a,b,c) = (0.15,0.2,10$) and funnel attractors ($(a,b,c)=(0.2,0.2,7)$). Making use of the symmetry of the maps with respect the main diagonal, the results concerning the phase-coherent case are depicted in the upper triangles, whereas results regarding oscillators with funnel attractors are shown in the lower one. For all 
simulations regarding system Eq.~\ref{eq:rossler_x_coupling}, the long time 
behavior of these quantities is calculated by averaging data between $t \in 
[4700,5000]$. One notices the formation of a $\pi$-state in the region 
with $r \simeq 0 $ (Fig.~\ref{Fig3}(a)) and $\delta \simeq \pi$ (Fig.~\ref{Fig3}(b))  
which is surrounded by areas of partial synchronization with $0 < \delta < \pi$ for both types of attractor, yet less prominent for funnel (notice that $\delta$ assumes random values in the incoherent region). 
Interestingly, in the dynamics of phase oscillators under attractive and repulsive interactions, one 
would expect 
the emergence of a collective rhythm different from the natural frequency in the lag-sync region~\cite{HoStr11PRL,IaPeMcSte13,sonnenschein2015collective}. 
Differently, weighted $x$-coupled R\"ossler oscillators subjected to weak noise intensities 
present $\Omega \simeq \omega_0$ for $0 \leq \delta \leq \pi$. On the other hand, for large negative values of $K_0G_0$, significant values of $\Omega \neq 
\omega_0$ are observed in the incoherent region, as shown in Fig.~\ref{Fig3}(c). Correspondingly, we refer to this regime as incoherent TW.

Comparing Fig.~\ref{Fig2} and Fig.~\ref{Fig3}(c) we see that $|\Omega - \omega_0|$ increases as one approaches the boundaries
of the escaping region, predicting in a way the occurrence of divergent trajectories.   
We stress though that this behavior is not a
result due to fluctuations, but an emergent phenomenon
that is yielded by the repulsive couplings considered in 
Eq.~\ref{eq:rossler_x_coupling}. Evidences of that can already be
seen in Fig.~\ref{Fig3}. Note that deviations from the natural
frequency $\omega_0$ are restricted to TW areas, while the $\pi$- and partial 
synchronized states are marked by an insignificant drift in the average frequency. In~\cite{SM} we further show that even without coupling mismatches in system 
(\ref{eq:rossler_x_coupling}), incoherent TWs emerge for large negative values of 
$K_0G_0$.

An evaluation of how 
noise impacts the formation of TW and $\pi$-states is shown in Fig.~\ref{Fig3}(d) considering phase-coherent oscillators.  Of particular interest is the non-monotonic dependence on the noise strength
by the $\pi$-state area, which, after a minimum, increases until the escaping region is reached. On the other hand, large noise completely extinguishes the TW area. Similar results are found for oscillators with funnel attractors (not shown).

\begin{figure}[!btp]
\begin{center}
\includegraphics[width=1.0\linewidth]{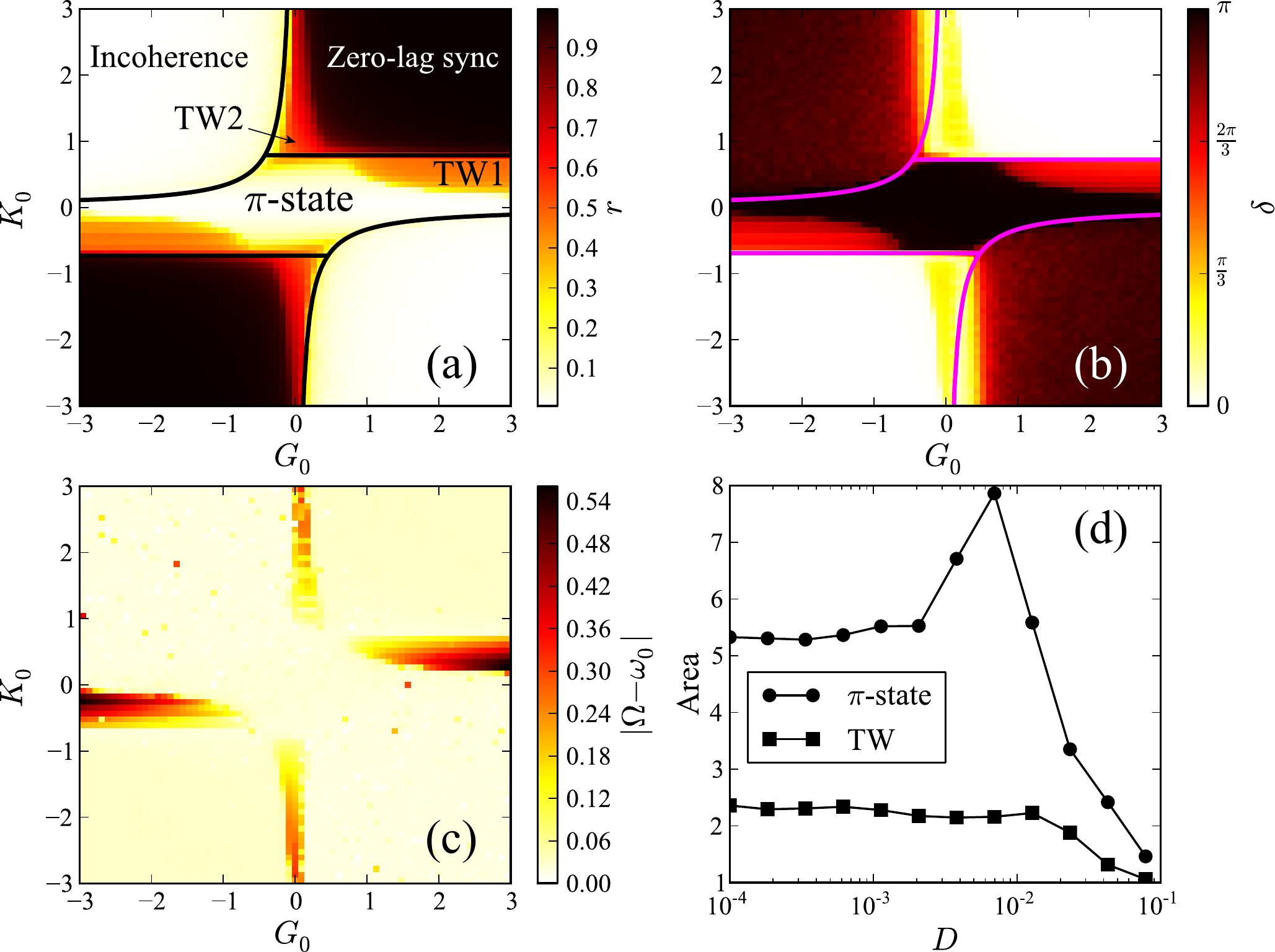}
\end{center}
\caption{Colormaps for the phase-coupled model (Eq.~\ref{eq:roessler}) showing (a) (total) order parameter $r$, (b)  phase lag $\delta$, (c) the spontaneous drift measured by $|\Omega - \omega_0|$, where $\Omega$ is the wave speed (Eq.~\ref{eq:wave_speed}) and $\omega_0$ the natural frequency. (d) Area of  TW state in which $|\Omega - \omega_0| \geq \varepsilon$, where we consider $\varepsilon = 0.3$; and $\pi$-state for $\delta \geq \varepsilon$ with $\varepsilon = 3$. For each point in the $70 \times 70$ grid, simulations are performed by evolving equations (\ref{eq:roessler}) considering initial conditions $\rho_i(0)$ and $z_i(0)$ randomly chosen from uniform distribution $[-1,1]$ and $\phi_i(0)$ uniformly distributed in the range $[-\pi,\pi]$. Other parameters: $N = 10^3$, $\Delta K = 1.45$, $\Delta G = 1$, $D = 0.03$, averaging over $t\in [4000,5000]$ and $dt = 0.05$.  Curves in the interface between incoherent and partial synchronized state are given by the condition $2D = \langle \langle KG \rangle \rangle$. Horizontal lines correspond to the critical points $K_{0c} = \pm \Delta K/2$.}
\label{Fig4}
\end{figure}

It is well-known that phase synchronization of weakly chaotic oscillators 
exhibits similar properties as in the dynamics of phase oscillators~\cite{rosenblum1996phase,*rosenblum1997FromPhaseLagSync,osipov2007synchronization}. However,
despite the striking contrast between the nature of the TW states observed in these systems, it is indeed possible to quantitatively describe certain aspects of the dynamics of chaotic oscillators with theories developed in the context of pure phase oscillators. In~\cite{sonnenschein2015collective} it is shown that the incoherent state ($r=0$) loses its stability if $2D = \langle \langle KG \rangle \rangle$ is satisfied, where 
$\langle \langle ... \rangle \rangle = \int dK' \int dG' ... P(K',G')$ with $P(K',G')$ being the joint probability distribution of in- and out-couplings. Applying this condition in the weak noise case and considering the effective noise strength $D_{\textrm{eff}}$~\cite{Deff},  we see that the result uncovered
for the Kuramoto model predicts with great accuracy the boundaries of incoherent states, reinforcing the idea that in regimes of weak noise and coupling the dynamics of chaotic oscillators is akin to periodic oscillators even though the amplitudes evolve chaotically and generally uncorrelated. 

Another critical condition is found by noticing that at $K_{0,c} = \pm \Delta K/2$ the oscillators of one of the populations are completely decoupled from the network, while the remaining connected oscillators are in a partially synchronized state. These conditions are depicted by the horizontal lines in Figs.~\ref{Fig3}(a) and (b), which encompass the $\pi$-state area.

\textit{Phase-coupled R\"ossler systems.}--- The levels of synchronization uncovered in 
the weighted $x$-coupled model share great similarity with the ones found in ensembles of 
phase oscillators subjected to attractive and repulsive couplings~\cite{HoStr11PRL,IaPeMcSte13,sonnenschein2015collective}. However, careful 
inspection of the collective frequency diagrams (Fig.~\ref{Fig3}(c)) reveals that the dynamical states in these systems are in fact different. Precisely, TWs among weighted $x$-coupled chaotic oscillators 
are only manifested in the absence of coherent oscillations, while partial synchronization 
is crucial for the emergence of such states in coupled phase oscillators. This poses the question of whether incoherent TWs are induced by the intrinsic chaotic dynamics or by the particular type of coupling adopted in Eq.~\ref{eq:rossler_x_coupling}. 

To shed light on this, we consider now a system of $N$ R\"ossler systems with Kuramoto phase-couplings~\cite{Kur84}.
In the phase-coherent regime it is valid to separate the original system into phase and amplitude dynamics. For this purpose, one goes to cylindrical coordinates $(\rho,\phi,z)$ via $x(t)=\rho(t)\cos\phi(t)$ and $y(t)=\rho(t)\sin\phi(t)$. This results in
$
\dot\rho =\ a\rho\sin^2\phi-z\cos\phi,\ \dot\phi =\ \omega + a\sin\phi\cos\phi + z/\rho \sin\phi,\ \dot z =\ b + \rho z \cos\phi - cz. 
$
We then formulate the phase-coupled model as
\begin{equation}
\begin{aligned}
\dot{\phi}_{i} = &\ \omega_{0}+a\sin\phi_{i}\cos\phi_{i}+\frac{z_{i}}{\rho_{i}}\sin\phi_{i}\\
 &\ +\frac{K_{i}}{N}\sum_{j=1}^{N}G_{j}\sin(\phi_{j}-\phi_{i})+\xi_{i}(t),\\
\dot{\rho}_{i} = &\ a\rho_{i}\sin^{2}\phi_{i}-z_{i}\cos\phi_{i},\\
\dot{z}_{i} = &\ b+\rho_{i}z_{i}\cos\phi_{i}-cz_{i},
\label{eq:roessler}
\end{aligned}
\end{equation}
$i=1,...,N$. We consider here $a = 0.15$, $b = 0.2$ and $c = 10$ so that the oscillators are in the phase-coherent regime. The parametrization of the coupling strengths is again adopted as in Eq.~\ref{eq:parametrization}.

We continue our analysis by getting a general view 
of the possible dynamical states of system (\ref{eq:roessler}). In order to do so, 
Fig.~\ref{Fig4} shows the simulation results 
of the total order parameter $r$, phase lag $\delta$ between the two populations and the wave speed $\Omega$, which are now calculated with respect to the phases defined in Eq.~\ref{eq:roessler}. For each of the $70 \times 70 $ points in the grid, system (\ref{eq:roessler}) is numerically integrated by using the Heun's scheme with time step $dt = 0.05$ and considering a population of $N = 10^3$ oscillators for  
which the quantities of interest are averaged over $t 
\in [4000,5000]$. By inspecting 
panels (a)-(c) of Fig.~\ref{Fig4}, three well-defined states 
are uncovered, namely incoherent ($r_{1,2} \simeq 0 $, $r \simeq 0 $ and $|\Omega - \omega_0| \simeq 0$), zero-lag sync 
($r \simeq 1 $, $\delta \simeq 0$ and $\Omega \simeq 
0$), $\pi$-state ($r \simeq 0 $, $\delta \simeq \pi$ and 
$|\Omega - \omega_0| \simeq 0$) and TW ($r > 0  $, $0< \delta < \pi$ and 
$\Omega \neq \omega_0$). Motivated by the 
findings in~\cite{sonnenschein2015collective}, we depict the two TW regions with 
distinct labels (TW1 and TW2) in order to highlight the 
different routes to these states. Specifically, TW1 is 
immersed in $\pi$-state region, whereas TW2 is surrounded 
by zero-lag sync (see also Fig. S2 in~\cite{SM}).
Furthermore, it is interesting to note that, in contrast to the weighted $x$-coupled model, 
TWs are not observed in the presence of incoherence if the oscillators are coupled through the phases. Another difference between the coupling models is the fact that the areas of the $\pi$-state behave in a slightly different manner as a function of the noise strength $D$ (compare panels (d) of Fig.~\ref{Fig3} and~\ref{Fig4}).

\textit{Conclusion.}--- 
The coexistence of chaotic, noisy and regular collective dynamics can be observed in many real-world systems~\cite{glass2001synchronization}. Here we provided the first evidences that TWs, i.e states in which a new rhythm among the oscillators different from the natural frequency emerges, are also attainable 
in the phase synchronization of chaotic R\"ossler oscillators. Two coupling formulations were analyzed. In the 
first, oscillators are connected via weighted linear coupling in the $x$ coordinates. Such interaction was shown to give rise 
to the novel state of incoherent TWs characterized by the appearance of new rhythm of oscillation in the absence of global 
synchronization. This suggests to employ the detection of additional frequency drifts to unveil new collective states in more general coupling schemes'.

Considering Kuramoto phase couplings, TWs are no longer found together with  incoherence, 
but rather only in regions with $0 < \delta <\pi$, consisting in the identical routes to such states as found 
in~\cite{sonnenschein2015collective}, despite the existence of chaotic amplitude dynamics.  

Our findings are of great general interest, because on the one hand, they reinforce the potential that phase models possess in 
describing the dynamics of higher dimensional systems, but on the other hand also highlights its limitations regarding other kinds of couplings, such as in the prediction of incoherent TWs. It remains as a future research to thoroughly determine the boundaries of the escaping region in Fig.~\ref{Fig2} as well as the investigation of whether other chaotic systems also exhibit similar dynamical patterns as the 
ones described here. Likewise promising is the investigation of multistable chimeralike states~\cite{omelchenko2011loss} in the context of attractive and repulsive couplings. Finally, we further expect to observe the aforementioned states in real experiments with, for 
instance, electrochemical and Belousov-Zhabotinsky 
oscillators~\cite{kiss2007engineering,*taylor2011phase,*totz2015phase} coupled via asymmetric mixed interactions. 
 
\acknowledgments 

T.K.DM.P. acknowledges FAPESP (grant 2012/22160-7 and 2015/02486-3) and IRTG 1740.  J.K. acknowledges IRTG 1740 (DFG and FAPESP). F.A.R. acknowledges CNPq (grant 305940/2010-4), 
FAPESP (grant 2013/26416-9) and IRTG 1740. L.SG. acknowledges support of Humboldt-University at Berlin within the framework of German excellence initiative (DFG). B.S. acknowledges funding from the Bundesministerium f\"ur Bildung und Forschung (BMBF) (BCCN II A3, grant 31401211). We thank A. B. Neiman, D. Eroglu, P. Schultz, M. Mungan and T. Pereira for stimulating discussions.

\bibliography{bibliography}

\clearpage


\section*{Supplementary material}

# Supplemental Material: Traveling phase waves in asymmetric networks of noisy chaotic attractors

Thomas K. DM. Peron[1,2],[*] Jürgen Kurths[2,3], Francisco A. Rodrigues[4], Lutz Schimansky-Geier[3], and Bernard Sonnenschein[3][†]

[1]*Instituto de Física de São Carlos, Universidade de São Paulo, CP 369, 13560-970 São Carlos, São Paulo, Brazil*
[2]*Potsdam Institute for Climate Impact Research (PIK), 14473 Potsdam, Germany*
[3]*Department of Physics, Humboldt-Universität zu Berlin, Newtonstrasse 15, 12489 Berlin, Germany and*
[4]*Instituto de Ciências Matemáticas e de Computação, Universidade de São Paulo, CP 668, 13560-970 São Carlos, São Paulo, Brazil*

## I. TRAVELING WAVES IN THE ABSENCE OF COUPLING MISMATCH ($\Delta K = \Delta G = 0$)

In the main text, it was shown that incoherent TWs emerge for large negative values of $K_0 G_0$ in the presence of in- and out-coupling mismatches. However, this poses the question about whether such states are also attainable in the absence of coupling mismatches between the oscillators. Figure S1 shows the order parameter $r$ and the drift in the frequencies $|\Omega - \omega_0|$ of weighted $x$-coupled Rössler oscillators (Eq. 1 in the main text) with phase-coherent and funnel attractors as function of $K_0$ and $G_0$ considering $\Delta K = \Delta G = 0$. Again, as in the case where $\Delta K \neq 0, \Delta G \neq 0$, TWs are found in regions with sufficiently negative values of $K_0 G_0$. Interestingly, this shows that the presence of a phase-lag between two populations is not a necessary condition for the emergence of a spontaneous drift different from the ensemble's mean frequency, contrary to what is observed in the collective dynamics of conformists and contrarians phase oscillators [S1, S2]. In fact, as shown in [S2], not only the condition $\delta \neq 0$, but also asymmetric coupling is required for the existence of TWs in ensembles of phase oscillators. This suggests that asymmetric coupling is not a necessary condition neither for the observation of incoherent TWs.

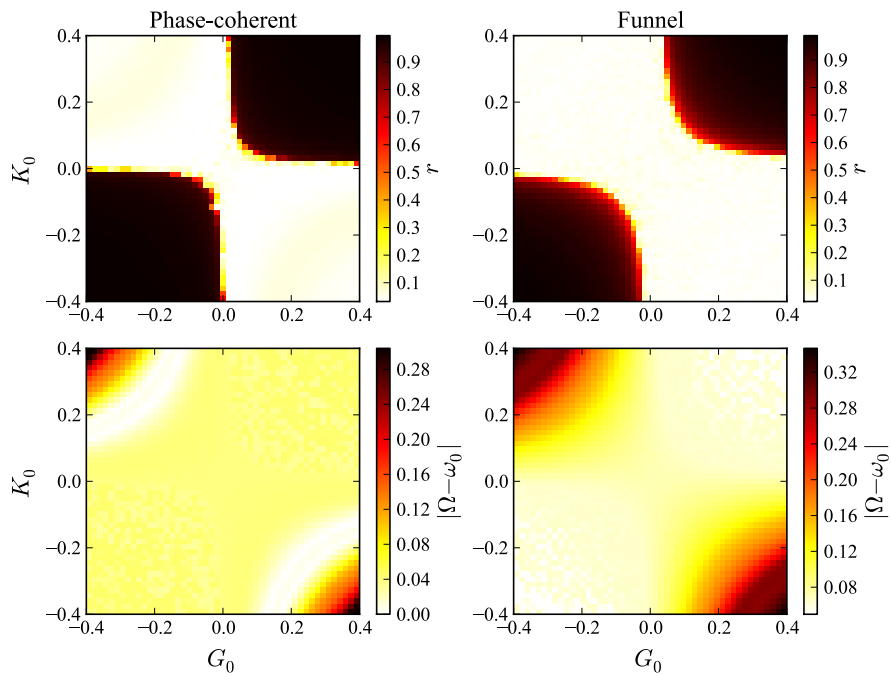

FIG. S1. Colormaps for weighted $x$-coupled Rössler oscillators with phase-coherent ($a = 0.15$, $b = 0.2$ and $c = 10$) and funnel attractors ($a = 0.2$, $b = 0.2$ and $c = 7$) in the absence of coupling mismatches, i.e. $\Delta K = 0$, $\Delta G = 0$. Parameters: $N = 10^3$, $dt = 0.05$. For each $(G_0, K_0)$ in the $50 \times 50$ grid, the quantities are averaged over $t \in [4500, 5000]$. Initial conditions $(x_i(0), y_i(0), z_i(0))$ are selected from an uniform distribution in the range $[-1, 1]$.

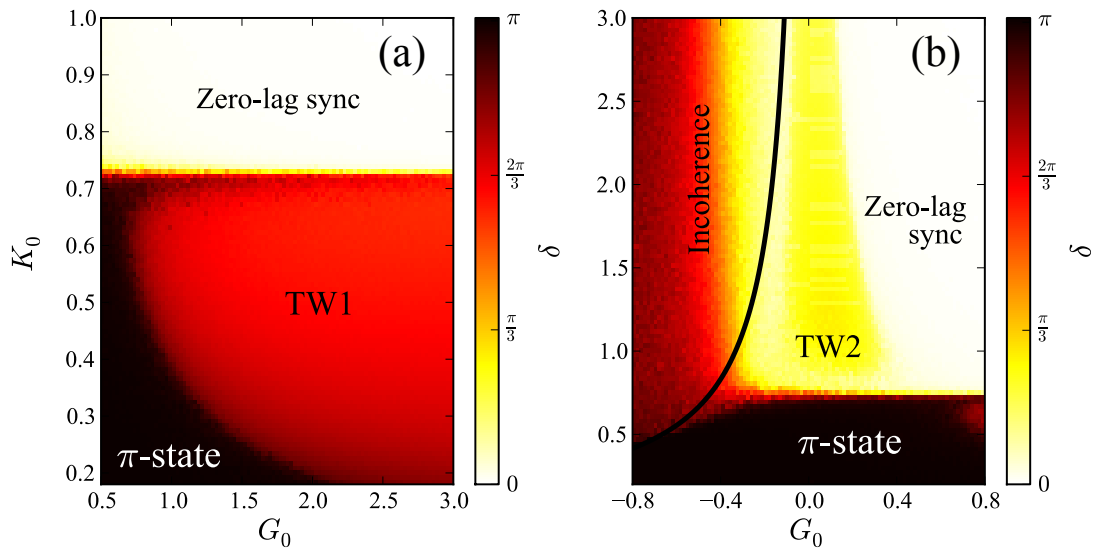

FIG. S2. Details of the (a) TW1 and (b) TW2 regions in Fig. 1 of the main text. Parameters: $N = 10^3$, $\Delta K = 1.45$, $\Delta G = 1$, $D = 0.03$, $dt = 0.05$. For each $(G_0, K_0)$ in the $100 \times 100$ grid, the quantities averaged over $t \in [4000, 5000]$. Solid black line in panel (a) is obtained by the critical condition $2D = \langle\langle KG \rangle\rangle$, where $\langle\langle ... \rangle\rangle = \int dK' \int dG' ... P(K', G')$ and $P(K, G)$ the joint probability distribution of in- and out-couplings.

Note that here only a single coupling strength exists, namely $K_0 G_0$. However, in order to better compare with the main text, we show the results in the $G_0 \times K_0$ plane.

## II. ADDITIONAL DIAGRAMS FOR THE PHASE-COUPLED MODEL

Figure S2 shows zoomed colormaps around the TW regions of Fig. 4 in the main text related to the phase-coupled model (Eq. 5 of the main text). We distinguish the TW states in panels (a) and (b) by TW1 and TW2, respectively, in order to highlight the different routes leading to each state. Precisely, the TW1 state corresponds to the TW state surrounded by the region $\pi$-state (Fig. S2 (a)), whereas TW2 stands for the TW state reached via the zero-lag state (Fig. S2 (b)). We recall that the emergence of TW and $\pi$-states as observed in Fig. S2 in the phase-coupled Rössler model bears an interesting similarity with the stochastic Kuramoto model in [S2], despite the fact that in the former the phases evolution explicitly depend on the chaotic amplitudes.



\end{document}